\documentstyle[12pt]{article}
\begin{document}
\hskip 9.5 truecm
IFUP-TH 19/94
\vspace{.5cm}
\begin{center}{\Large Massless scalar fields in 1+1 dimensions
and Krein spaces}\\
\vspace{.7cm}
A. Z. Capri\\
Department of Physics and Theoretical Physics Institute \\
University of Alberta, Edmonton, Alberta, T6G 2J1, Canada\\
\vspace{.5cm}
R. Ferrari\\ Dipartimento di Fisica, Universit\`a di Trento\\
38050 Povo, Trento, Italy \\
and\\
I.N.F.N., Sezione di Padova\\
\vspace{.5cm}
V. Montalbano \\
Dipartimento di Fisica,
Universit\`a di Pisa \\
56100 Pisa, Italy \\
and \\
I.N.F.N., Sezione di Pisa
\end{center}
\vspace{1cm}
\begin{abstract}
We consider the Krein realization of the Hilbert space for a
massless scalar field in 1+1 dimensions.
We find convergence criteria and the completion of the space of
test functions ${\cal S}$ with the topology induced by the Krein
scalar product.
Finally, we show that the interpretation for the Fourier components as
probability amplitudes for the momentum operator is lost in this case.
\end{abstract}
\vskip .5 truecm
\par\noindent
11.10 Cd
\newpage
\section{Introduction}
\label{sec:int}
The massless scalar field in $1+1$ dimensions is affected by the
problem
of indefiniteness of the metric \cite{schoer}.
One would like to represent the
algebra of the scalar massless field on a Hilbert space
by staying as close as possible to Wightman's axioms. However,
if the metric is not positive
definite, one can not construct  the Hilbert
space by completing the vector space obtained from the smeared
field.
\par
In this note we would like to focus on some of the difficulties
one encounters in defining this apparently simple field theory.
\par
Let us assume that
a vacuum  exists such that
\begin{equation}
\langle 0 | \phi(f) | 0 \rangle = 0
\label{vev}
\end{equation}
and
\begin{eqnarray}
&&\langle g,f\rangle \equiv
\langle 0|\phi(g)\phi(f) |0 \rangle =
\nonumber \\
&&{1\over {4\pi}}\int_{\cal R}
{{dp}\over{|p|}} \left [{ g}^*(|p|,p)
{ f}(|p|,p)    -
{ g}^*(0,0){ f}(0,0) \theta(1-|p|)\right ],
\label{tpf}
\end{eqnarray}
where $f,g \in {\cal S}({\cal R}^2)$. The two-point function
gives the sesquilinear form, which in a good theory is the
scalar product. Here, due to the subtraction, the induced metric
is not positive definite.
\par
This goal is not only a challenging problem: it is also related
to some important issues of field theory. One is whether
the negative norm allows an evasion of Coleman's theorem
and thus there is a spontaneous breakdown of the symmetry
\begin{equation}
\phi(g)\to \phi(g) + {\rm constant}\cdot g(0)
\label{sym}
\end{equation}
(solid state physics has an equivalent theorem by Mermin and
Wagner). Moreover during the discussions a few surprising
facts appear. We hope that they are not paradigms
for more realistic cases.
\par
It has been suggested \cite{dubin,strocchi1} that a natural
solution of this problem is in the use of the Krein metric
\cite{krein}.
We have reconsidered this problem and reached conclusions that
are  somewhat different from those of previous authors.
We find that Stone's theorem is evaded for the
translation operator. In particular one looses the
interpretation for the Fourier components as probability
amplitudes for the momentum operator.
Our conclusion is that one needs a new strategy.
\section{Krein space}
\label{sec:kre}
In this sections we consider the vector space $\cal F$
of functions $f(|p|,p), f\in
{\cal S}({\cal R}^2)$ where a sesquilinear form is given
in eq. (\ref{tpf}). Please notice that continuous functions
cannot be used (counter example: take a function which behaves
like $(\ln |p|)^{-1}$  near
the origin, then the integral
does not exists. The reason is that the sesquilinear form
is a well defined distribution, but not a Radon measure.
This fact is the origin of all the troubles which we will
find).
\par
Let us recall the definition of  Krein space \cite{krein}.
A vector space ${\cal F}$ with a sesquilinear Hermitian form Q
defined on it is said to have a Q-metric
\begin{equation}
 Q(f,g) \equiv \langle f,g\rangle \qquad f,g \in {\cal F},
\end{equation}
in general indefinite.
\par
\noindent
If this space furthermore admits a canonical decomposition
\begin{equation}
{\cal F} = {\cal F}^{(+)}\ [\dot+]\ {\cal F}^{(-)}
\label{candec}
\end{equation}
where ${\cal F}^{(+)}$ is orthogonal in the Q-metric to ${\cal
F}^{(-)}$
and both ${\cal F}^{(+)}$ and ${\cal F}^{(-)}$ are Hilbert spaces
with
norms
\begin{eqnarray}
\|f\|^2 \equiv \langle f,f\rangle \qquad f \in {\cal
F}^{(+)}\nonumber\\
\|f\|^2 \equiv  - \langle f,f\rangle \qquad f \in {\cal F}^{(-)}
\end{eqnarray}
then  ${\cal F}$ is a Krein space.
\par
\vspace{ .5 cm}
\centerline{{\bf Canonical decomposition}}
\par
This section is devoted to finding the canonical decomposition of eq.
(\ref{candec}). We state a few simple lemmas:
\par\noindent
Lemma 1
\par
{\it If $f(0) = 0 \quad \Rightarrow \quad \langle f,f\rangle \geq 0$,
where the case $\langle f,f\rangle = 0$ implies $f = 0$.}
\par\noindent
{}From this follows
\par\noindent
Lemma 2
\par
{\it If  $f \not= 0 \quad and \quad\langle f,f\rangle =
0\quad \Rightarrow f(0) \not= 0$}.
\par\noindent
Lemma 3
\par
{\it
 Let $f$ and $g$  be any two linearly independent functions.
If $\langle f,f\rangle \leq 0$ and $\langle g,g\rangle \leq 0 $,
then $\langle f,g\rangle \not= 0$.}
\par\noindent
${\underline{\rm Proof}}$. Since $f \not= 0$ and $g \not= 0$ and
their sesquilinear form is not
positive,  we get form Lemma 1 $f(0) \not= 0$ and $g(0) \not= 0$.
Define
\begin{equation}
h \equiv g(0)f - f(0)g
\end{equation}
then $h(0) =0$ but $h\not =0$. {}From Lemma 1 it follows
that $\langle h,h\rangle >0$. This conclusion is compatible only
with $\langle f,g\rangle\not =0$, otherwise
\begin{equation}
\langle h,h\rangle = |g(0)|^2\langle f,f\rangle +
|f(0)|^2\langle g,g\rangle\leq 0.
\end{equation}
\par
Now we can prove the following theorem (see Ref. \cite{dubin})
\par\noindent
Theorem 1
\par
{\it The vector space admits  a canonical decomposition }
\begin{equation}
{\cal F} = {\cal F}^{(+)} [\dot+] {\cal F}^{(-)}.
\end{equation}
and {\it  ${\cal F}^{(-)}$  contains only one element}.
\par\noindent
${\underline{\rm Proof}}$. First we prove that every vector
subspace ${\cal F}^{(-)}$ contains only one linear independent
element. In fact suppose that
it contains more than one. Let $f, g \in {\cal F}^{(-)}$.
Then from Lemma 1 we get
$f(0)\not= 0$ and $g(0)\not= 0$. We define
\begin{equation}
h \equiv f(0)g - g(0)f.
\end{equation}
Since $h(0)=0$, it follows that  $\langle h,h\rangle > 0$ and
therefore $h\not\in {\cal F}^{(-)}$.
\par
Consider an element of $\chi\in{\cal F}$ with
\begin{equation}
\langle\chi,\chi\rangle = -1
\end{equation}
and define
\begin{eqnarray}
f_+& =& f + \langle\chi,f\rangle\chi
\nonumber \\
f_-& =& - \langle\chi,f\rangle\chi.
\label{ca}
\end{eqnarray}
Then, by construction,
\begin{equation}
\langle f_+,f_-\rangle = 0.
\end{equation}
\par\noindent
Lemma 4
\par\indent
$\langle f_+,f_+\rangle \geq 0$ ($\langle f_+,f_+\rangle = 0$
only if $f_+ = 0$).
\par\noindent
${\underline{\rm Proof}}$. Since $\langle\chi,f_+\rangle = 0$
and $\langle\chi,\chi\rangle = -1$ it cannot be that
$\langle f_+,f_+\rangle \leq 0$. In fact if
$\langle f_+,f_+\rangle \leq 0$ one can apply Lemma 2 and get
$\langle f_+,\chi\rangle \not= 0$ which is contrary to the
assumption.
\par
We prove now a kind of anti-Schwartz inequality, which is valid
for any pair of functions with sesquilinear form less than or equal
to  zero
(consequently they {\bf have} to be  non-zero in zero,
otherwise the norm is strictly positive).
\par\noindent
Lemma 5
\par\indent
{\it Let $f$ and $g$ be two non-zero linearly independent continuous
functions with non-positive sesquilinear form:}
\begin{equation}
\langle f,f\rangle \leq 0\quad {\rm and}\quad \langle g,g\rangle \leq
0
\end{equation}
{\it then}
\begin{equation}
|\langle f,g\rangle|^2 > \langle f,f\rangle\langle g,g\rangle.
\end{equation}
${\underline{\rm Proof}}$.
Let us first consider the case
\begin{equation}
\langle f,f\rangle = \langle g,g\rangle = 0.
\end{equation}
Define
\begin{equation}
h \equiv f(0)g - g(0)f .
\end{equation}
Since, by construction $h(0) = 0$, but $h \not= 0$ ($f$ and $g$ are
linearly
independent), then
\begin{equation}
\langle h,h\rangle > 0
\end{equation}
and therefore
\begin{equation}
- f^*(0)g(0)\langle g,f\rangle - f(0)g^*(0)\langle f,g\rangle > 0.
\end{equation}
This is possible only if
\begin{equation}
|\langle f,g\rangle|^2 > 0.
\end{equation}
Now let $\langle g,g\rangle < 0$. We define
\begin{equation}
k \equiv f - g {\langle g,f\rangle \over \langle g,g\rangle}.
\end{equation}
By construction
\begin{equation}
\langle g,k\rangle = 0\quad {\rm and}\quad k \not= 0.
\end{equation}
This is possible only if $\langle k,k\rangle > 0$ (see
Lemma 2). Therefore
\begin{equation}
\langle k,k\rangle = \langle f,f\rangle  + {|\langle g,f\rangle|^2
\over
\langle g,g\rangle} - 2{|\langle f,g\rangle|^2 \over \langle
g,g\rangle} > 0.
\end{equation}
Finally
\begin{equation}
|\langle f,g\rangle|^2 > \langle f,f\rangle\langle g,g\rangle .
\end{equation}
\par
\vspace{ .5 cm}
\centerline{{\bf Krein metric}}
\par
We can now introduce a positive metric (Krein metric) by using
the canonical decomposition provided in eq. (\ref{ca}).
\begin{equation}
( f, g) \equiv \langle f_+,g_+\rangle + \langle f,\chi\rangle
\langle \chi,g\rangle.
\label{k}
\end{equation}
Notice the following alternative form. Use
\begin{equation}
f_+ = f +\langle\chi,f\rangle\chi
\end{equation}
then
\begin{equation}
\langle f_+,g_+\rangle = \langle f,g\rangle +
\langle\chi,g\rangle\langle f,\chi
\rangle
\end{equation}
and finally
\begin{equation}
( f, g ) = \langle f,g\rangle + 2\langle f,\chi\rangle
\langle \chi,g\rangle.
\label{kp}
\end{equation}
\par
{}From the last two equations we have important inequalities:
\begin{equation}
\langle f,f\rangle \geq - \langle f,\chi\rangle\langle \chi,f\rangle
\end{equation}
and
\begin{equation}
\langle f,f\rangle \geq - 2\langle f,\chi\rangle\langle
\chi,f\rangle.
\end{equation}
The norm induced by Krein's scalar product also satisfies the
triangle
inequality.
\par
The positive metric introduced in Ref. \cite{strocchi1} differs
from ours. See  \cite{strocchi2}.
\section{Convergence criteria}
The problem is now to consider all possible Cauchy sequences in the
Krein norm. We prove in this section that the Krein metric is not
necessary for this goal. Let us introduce the notion of strong
and weak convergence associated to the sesquilinear form:
\par\noindent
${\underline{\rm Weak~~ convergence}}$: $f_n$ is said to converge
weakly to $f$ if for any $g$
\begin{equation}
\lim_{n\to\infty} \langle g,
(f_n-f)\rangle  = 0
\end{equation}
\par\noindent
${\underline{\rm Strong~~ convergence}}$. $f_n$ is said to converge
strongly to $f$ if
\begin{equation}
\lim_{n\to\infty} \langle (f_n-f),
(f_n-f)\rangle  = 0.
\end{equation}
\par
The surprising result is:
\par\noindent
Theorem 2
\par
{\it  Convergence in the Krein metric
is equivalent to convergence  in
the sesquilinear form (if both weak and strong).}
\par\noindent
${\underline{\rm Proof}}$. 1) Let $f_n$ a Cauchy sequence in the
Krein metric. We use the relation
\begin{equation}
\langle f,g\rangle = (f,g) - 2\langle\chi,g\rangle\langle
f,\chi\rangle.
\label{so}
\end{equation}
For $g=\chi$ we get
\begin{equation}
- \langle f,\chi\rangle = (f,\chi)
\end{equation}
and therefore by using the Schwartz inequality for the positive
scalar
product $(.,.)$ we find
\begin{equation}
|\langle f,\chi\rangle| = |(f,\chi)| \leq \Vert f\Vert.
\end{equation}
Finally from eq. (\ref{so}) we get
\begin{equation}
|\langle f,g\rangle| \leq |(f,g)| + 2|\langle\chi,g\rangle||\langle
f,\chi
\rangle| \leq 3\Vert f\Vert\Vert g\Vert.
\end{equation}
The last equation implies both weak and strong convergence
in the sesquilinear form.
\par\noindent
2) Let $f_n$ be a Cauchy sequence in the weak and strong sense in the
sesquilinear form. We use eq. (\ref{so})
\begin{eqnarray}
([f_n-f_{n'}],[f_n-f_{n'}]) &=&
\langle [f_n-f_{n'}],[f_n-f_{n'}]\rangle +
2|\langle\chi,[f_n-f_{n'}] \rangle|^2
\nonumber \\
&\leq&
|\langle [f_n-f_{n'}],[f_n-f_{n'}]\rangle| +
2|\langle\chi,[f_n-f_{n'}] \rangle|^2
\nonumber \\
\end{eqnarray}
Then $f_n$ is a Cauchy sequence also in the Krein metric.
\par
To conclude the section we stress that the completion of $\cal F$
can be performed by using the indefinite metric.
\section{Hilbert space}
Now we consider all the Cauchy sequences in order to complete
the space. It is useful to choose
\begin{equation}
\chi(0) =1
\end{equation}
and to decompose
\begin{equation}
{\hat f} \equiv f-f(0)\chi.
\label{d}
\end{equation}
In Ref. \cite{strocchi1} a sequence is proposed with very
interesting properties. Here we use a similar one (its
support is in $p_0 >0$)
\begin{equation}
v_n = {{4\pi}\over{\ln n}}
{\tilde \theta}(-1 + 2np_0)
\chi(p)
\end{equation}
where
\begin{eqnarray}
&&{\tilde \theta}\in {\cal S}
\nonumber \\
&& 0\leq{\tilde \theta}\leq 1
\nonumber \\
&&{\tilde \theta}(x) = 0 \quad \forall  x\leq 0
\nonumber \\
&&{\tilde \theta}(x) = 1 \quad \forall  x\geq 1.
\end{eqnarray}
We denote by the same symbol $v_n$ the element of ${\cal F}$
given by $v_n(|p|,p)$.
\par\noindent
Lemma 6
\par
{\it  $v_n$ is a Cauchy sequence in ${\cal F}$.}
\par\noindent
Moreover
\par\noindent
Lemma 7
\par
{\it For any $f\in {\cal F}$}
\begin{equation}
\lim_{n\to\infty}\langle v_n, f\rangle = f(0).
\label{vf}
\end{equation}
Let us denote by $v$ the formal element of the Hilbert space
\begin{equation}
v \equiv \lim_{n\to\infty} v_n.
\label{v}
\end{equation}
The existence of the Cauchy sequence $v_n$ allows us to prove
\par\noindent
Lemma 8
\par
{\it If $\{ f_n = {\hat f}_n + f_n(0)\chi \}$ is a Cauchy sequence
then $f_0 \equiv \lim_{n\to\infty} f_n(0)$ exists}
\par\noindent
${\underline{\rm Proof}}$. Since both $f_n$ and $v_n$ are Cauchy
sequences then also
$\langle v_n, f_n\rangle = f_n(0)$ converges.
\par
Finally we have to characterize the limit of ${\hat f}_n$.
Notice that
\begin{equation}
\langle {\hat f}_n, {\hat f}_n\rangle =
{1\over {4\pi}}\int_{\cal R}
{{dp}\over{|p|}} |{\hat f}_n|^2(p).
\end{equation}
Therefore the limit of ${\hat f}_n$ is any function in
$L_2(dp/p,{\cal R})$ for which the integral
\begin{equation}
{1\over {4\pi}}\int_{\cal R}
{{dp}\over{|p|}} \chi^*(p){\hat f}(p)
\end{equation}
exists, i.e. ${\hat f}\in L(dp/p,K)$ where $K$ is some compact
interval
containing the point $p=0$.
\par
No other state should exists with support at $p=0$ only, in fact
the elements of ${\cal F}$ are not differentiable at this point.
\section{Problems with the translations}
There is a disturbing feature with the state $v$ given in
eq. (\ref{v}). Notice that at every point
\begin{equation}
\lim_{n\to\infty} v_n(p) = 0.
\label{loc}
\end{equation}
Thus, it is disturbing to find that the limit of $v_n$ in the Krein
metric is non zero. Our uneasiness comes from the prejudice
that the Fourier component at $p$ is (proportional to) the
probability amplitude for a state to have momentum $p$. Thus
if all components are zero we expect the vector to be zero.
The paradox is resolved if we look once again at the
distribution which defines the metric in eq. (\ref{tpf}).
Notice that
\begin{equation}
\lim_{n\to\infty} v_n(p) = 0
\label{loco}
\end{equation}
in the norm
\begin{equation}
||f|| = \sup_{x\in K} |f|
\label{cn}
\end{equation}
where $K\subset {\cal R}$ is  compact. A Radon measure is
the dual space of continuous functions in ${\cal R}$
with the above norm \cite{treves}.
But eq. (\ref{vf}) tells us that the limit is non zero. Thus
we have proven the following theorem
\par\noindent
Theorem 3
{\it The distribution
\begin{equation}
{1\over {4\pi}}\int_{\cal R} {{dp}\over{|p|}}\left [
e^{ipx} - \theta(1-|p|) \right ]
\end{equation}
is not a Radon measure}.
\par
The theorem above excludes also the possibility of a difference
of positive measures. On the other hand we expect the
translation operator to be written in terms of projection
operators on states of definite momentum (Stone's theorem)
\begin{equation}
U(a)= \int_{\cal R} e^{ipa} dE(p)
\label{stone}
\end{equation}
and that therefore the Fourier transform of the two-point function
should be a measure. Stone's theorem is evaded here because
$U(a)$ is unitary only in the indefinite metric (in the Krein
metric it is not unitary, see eq. (\ref{kp})).
\par
Finally we recall that the state $v$ is invariant under translations
\cite{strocchi1}. This is an immediate consequence of
eq. (\ref{vf}). Moreover the procedure can be repeated
on the vector space of any number of particles and thus an
infinite number of states invariant under translations can
be constructed. In particular, since $v_n$ has support only
in the region $p_0> 0$, one can define
\begin{equation}
|v;k\rangle \equiv \lim_{n\to\infty}
{1\over {\sqrt{k!}}} \left [\phi(v_n)\right]^k
|0\rangle.
\label{many}
\end{equation}
For $k\not=0$, one gets easily
\begin{equation}
\langle v;k, v;k\rangle = 0
\label{ny}
\end{equation}
and their Krein norm (we have assumed $\chi(0)=1$)
\begin{equation}
(v;k,|v;k) = \left(2|\langle \chi,v\rangle|^{2}\right)^k
=\left(2|\chi(0)|^2\right)^{k} = 2^k.
\label{any}
\end{equation}
Finally one can construct (for any complex number $\alpha$)
\begin{equation}
|0_\alpha \rangle \equiv \lim_{N\to\infty}
\sum_{k=0}^N {{i\alpha^k}\over{\sqrt{k!}}}
|v;k\rangle.
\label{y}
\end{equation}
with the properties
\begin{eqnarray}
\langle 0_\beta, 0_\alpha \rangle &=& 1
\nonumber \\
(0_\beta,0_\alpha )& =& e^{2\beta^*\alpha}.
\label{mm}
\end{eqnarray}
Since they are invariant under translations,
we prefer to call them vacuums.


\begin{thebibliography}{99}
\bibitem{strocchi2} The positive metric in Ref. \cite{strocchi1} is
given
by means of a ${\cal S}$ function $\xi$ such that $\xi(0)=1$ and
$\langle \xi, \xi \rangle=0$. Moreover $f_0 \equiv f-f(0)\xi$. Thus
%
$\langle f,g\rangle =\langle f_0,g_0\rangle +
f^*(0)\langle \xi,g\rangle +
g(0) \langle f,\xi\rangle$.
%
A "Krein metric" is defined by
$(f,g)=\langle f_0,g_0\rangle + \langle f,\xi\rangle
\langle \xi, g\rangle +
f^*(0)g(0)$.
%
Now if $f\in {\cal F}^{(+)}$ then
$|f(0) - \langle f,\xi\rangle|^2 = 0$.
If $f\in {\cal F}^{(-)}$ then
$2\langle f_0,f_0\rangle + |f(0) + \langle f,\xi\rangle|^2 = 0$,
i.e. $f=0$. In conclusion: it is not a Krein metric.
\end{thebibliography}
\end{document}